\documentclass[aps,prl,reprint,letterpaper,amsmath,amssymb, superscriptaddress,longbibliography]{revtex4-1}
\usepackage{graphicx}

\begin{document}
\title{Radiofrequency driving of coherent electron spin dynamics in $n$-GaAs detected by Faraday rotation}

\author{V.~V.~Belykh}
\email[]{belykh@lebedev.ru}
\affiliation{P.N. Lebedev Physical Institute of the Russian Academy of Sciences, 119991 Moscow, Russia}
\affiliation{Experimentelle Physik 2, Technische Universit\"{a}t Dortmund, D-44221 Dortmund, Germany}
\author{D.~R.~Yakovlev}
\affiliation{Experimentelle Physik 2, Technische Universit\"{a}t Dortmund, D-44221 Dortmund, Germany}
\affiliation{Ioffe Institute, Russian Academy of Sciences, 194021 St. Petersburg, Russia}
\author{M.~Bayer}
\affiliation{Experimentelle Physik 2, Technische Universit\"{a}t Dortmund, D-44221 Dortmund, Germany}
\affiliation{Ioffe Institute, Russian Academy of Sciences, 194021 St. Petersburg, Russia}

\date{\today}
\begin{abstract}
We suggest a new pump-probe method for studying semiconductor spin dynamics based on pumping of carrier spins by a pulse of oscillating radiofrequency (rf) magnetic field and probing by measuring the Faraday rotation of a short laser pulse. We demonstrate this technique on $n$-GaAs and observe the onset and decay of coherent spin precession during and after the course of rf pulse excitation. We show that the rf field resonantly addresses the electron spins with Larmor frequencies close to that of the rf field. This opens the opportunity to determine the homogeneous spin coherence time $T_2$, that is inaccessible directly in standard all-optical pump-probe experiments.
\end{abstract}
\maketitle


\textit{Introduction.} When a magnetic field $\mathbf{B}$ is applied to an ensemble of electron spins, they all precess with the Larmor frequency $\mathbf{\omega_\text{L}} = g\mu_\text{B}\mathbf{B}/\hbar$, where $g$ is the electron $g$ factor, $\mu_\text{B}$ is the Bohr magneton, and $\hbar$ is the Planck constant. This ensemble precession is hardly detectable since the relative phases of the spins are random so that macroscopic spin polarization averages to zero. However, we can induce a common phase by applying a weak radiofrequency (rf) magnetic field oscillating with a frequency $\omega$ close to $\omega_\text{L}$. In analogy with a driven harmonic oscillator, the electron spins after some time will resume the frequency and phase of the rf field. As a result the in-phase spin precession motions form a macroscopic spin polarization that can be detected optically by measuring the Faraday/Kerr rotation of the linear polarization of a laser pulse which arrival time is synchronized with the rf field oscillation. By applying an rf field of finite duration and scanning the relative delay of the optical probe it is possible to measure the coherent electron spin dynamics.

The method of rf spin pumping and optical probing, developed in this work, is inspired by the similar all-optical pump-probe technique \cite{Baumberg1994,Zheludev1994,Dyakonov2017}, and in particular by the extended pump-probe method \cite{Belykh2016}, in which electron spins are oriented by circularly-polarized laser pulses. By contrast, the rf pump directly addresses the {\it resident} electron spins at their Larmor precession frequencies, which is likely advantageous compared to optical pumping at the much higher frequencies of the resonant electron transitions, unrelated to the spin precession. Furthermore, optical pumping strongly perturbs the system by photoexciting exciton complexes whose constituents subsequently contribute to the spin dynamics. The optical readout of the developed technique (without rf), on the other hand, is similar to spin-noise spectroscopy \cite{Aleksandrov1981,Crooker2004,Oestreich2005}, where the noise induced by random spin precessions at the frequency $\omega_\text{L}$ is detected. However, spin noise spectroscopy finally provides the frequency spectrum rather than the spin dynamics. rf spin-resonant excitation is used in electron paramagnetic resonance (EPR), particularly in pulsed EPR, from which one can obtain the relevant spin relaxation times \cite{Blume1958,Gordon1958,Schweiger2001}. 
However, EPR lacks the detection sensitivity of optical techniques and also addresses the entire sample while optical probing can provide spatial resolution. Optically detected magnetic resonance (ODMR) experiments can be performed with temporal resolution \cite{Dawson1979,Trifunac1980}. In these experiments, the effect of a high-frequency field on the photoluminescence (PL) of a sample is detected, which requires fast photodetectors for resolving the coherent spin precession. The precession was observed using Raman heterodyne detection of the magnetic resonance \cite{Mlynek1983,Wei1996}, which is hardly applicable to resident electrons in semiconductors.

In the technique developed here, the spin polarization is probed directly by Faraday (or Kerr) rotation using a short laser pulse (for high temporal resolution) leaving the system's spin state unperturbed, in contrast to the PL measurement in standard ODMR. Further, the technique does not suffer from dead-times, during which the dynamics is inaccessible, and its temporal resolution is determined by the jitter between the rf and optical pulses which can be better than 25~ps.

In this paper we demonstrate the principle of a spin resonant pump-probe method by measuring the electron spin dynamics in bulk $n$-GaAs. The results are well-described by a model based on the Bloch equations \cite{Bloch1946,Abragam1961}. We show that rf excitation addresses the electron spins resonantly within a $1/T_2$ frequency  interval around the Larmor frequency allowing direct measurement of the spin coherence time $T_2$ and advanced spin manipulation.

\begin{figure}
\includegraphics[width=1\columnwidth]{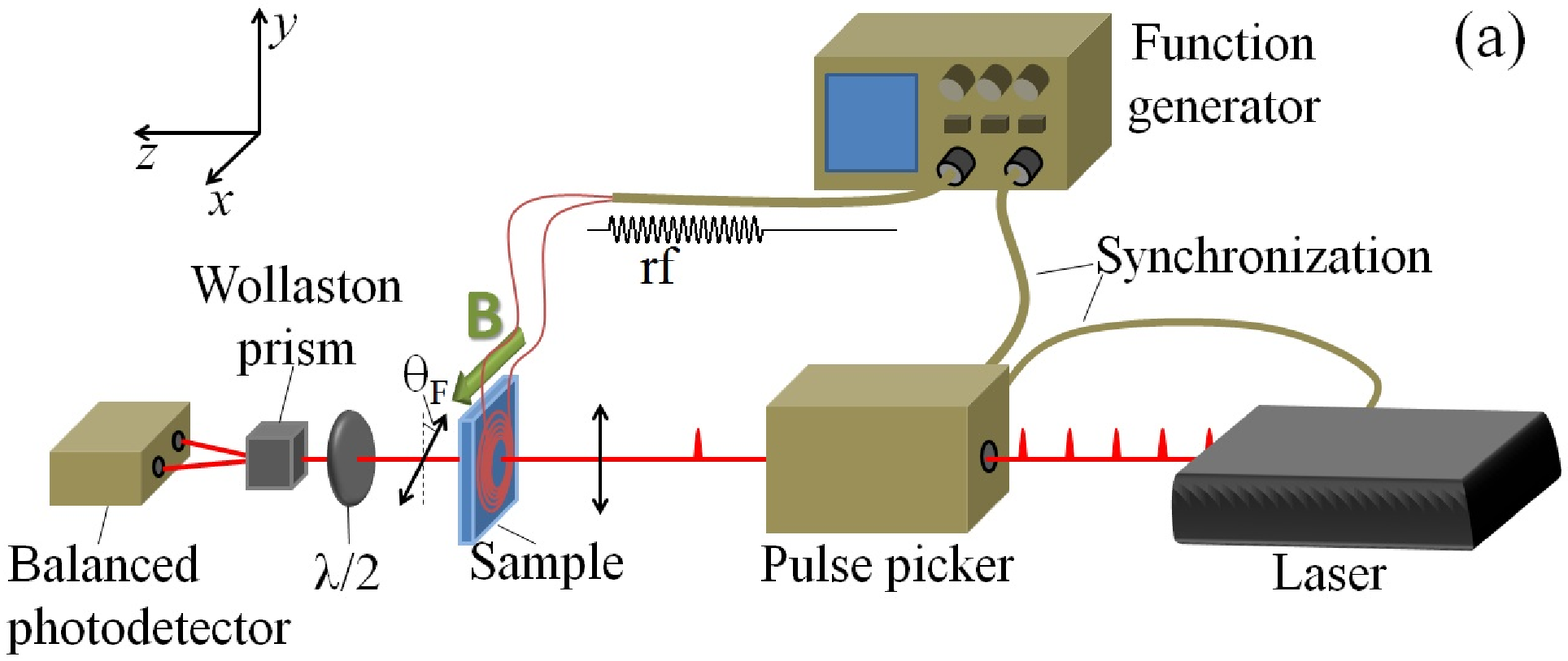}
\includegraphics[width=1\columnwidth]{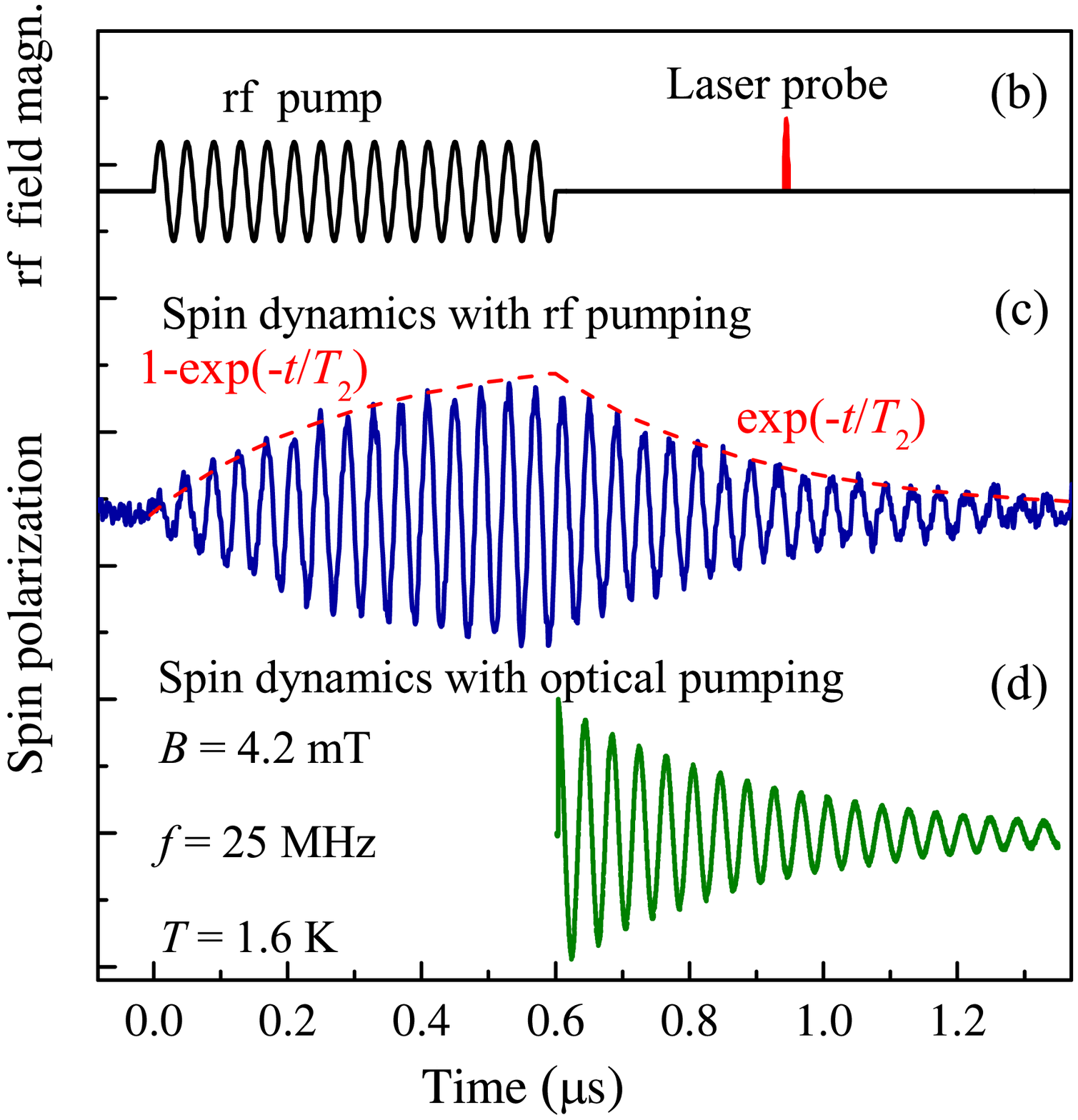}
\caption{(a) Scheme of experiment. The electron spins in the sample, precessing around the magnetic field $\mathbf{B}$, are oriented by rf field pulses generated by the function generator and applied via the coil. The spin polarization is probed by the Faraday rotation of the linear polarization of the optical pulses picked from the train of laser pulses. The dynamics of spin precession is scanned by changing the delay between the synchronized rf and optical pulses.
(b) Temporal profile of the rf field pulse applied to the sample. The repetition period of the rf pulses is 8.4~$\mu$s. (c) Dynamics of the electron spin polarization measured via probing the Faraday rotation during the course of rf pulse application. The dashed lines show calculated rising and decaying envelopes of the signal. (d) Measured dynamics of the electron spin polarization in an all-optical extended pump-probe Faraday rotation experiment (the pump pulse arrives at 0.6~$\mu$s). }
\label{fig:Scheme}
\end{figure}

\textit{Experimental details.} The experiment is performed on a Si-doped GaAs sample with electron concentration of $1.4 \times 10^{16}$~cm$^{-3}$ (350-$\mu$m-thick bulk wafer) slightly exceeding the metal-insulator transition. The sample is placed in the variable temperature ($T =1.6-300$~K) insert of a cryostat with a split-coil superconducting magnet, or in cryostat with permanent magnet placed outside on a controllable distance. A constant magnetic field $\mathbf{B}$ is applied perpendicular to the light propagation direction and to the sample normal (Voigt geometry).

The experimental scheme is presented in Fig.~\ref{fig:Scheme}(a). The rf magnetic field with an amplitude of $0.01-0.8$~mT is applied along the sample normal using a small ($\lesssim 1$~mm-inner and $\sim 3$~mm-outer diameter) coil near the sample surface. The current through the coil is driven by a function generator, which creates voltage pulses of sinusoidal form with a duration of 600~ns (unless stated otherwise). The oscillation frequency within each pulse is $f=25$~MHz (unless stated otherwise), which has to be compared to the Larmor precession frequency of the electron spins. The rf field pulse created by the coil acts as pump pulse. For probing we use 2-ps-long optical pulses generated by a Ti:Sapphire laser. The laser emits a train of pulses with a repetition rate of 76~MHz (repetition period $T_\text{R}=13.1$~ns), which is reduced to 238~kHz (repetition period $320 T_\text{R}=4.2$~$\mu$s) by selecting single pulses with an acousto-optical pulse picker synchronized with the laser. The arrival of the rf pulses is synchronized with the laser pulses (a signal of the pulse picker triggers the function generator) and the repetition period of the rf pulses is twice longer than that of the laser pulses. The latter is required for synchronous detection at the repetition frequency of the rf pulses of 119~kHz. The delay between the pump rf pulse and the probe laser pulse is changed electronically by the function generator. The probe laser pulses are linearly polarized and the Faraday rotation of their polarization after transmission through the sample is analyzed using a Wollaston prism, splitting the probe into two orthogonally polarized beams which are registered by a balanced photodetector. The laser wavelength is set to 825-830~nm.

\begin{figure}
\includegraphics[width=1\columnwidth]{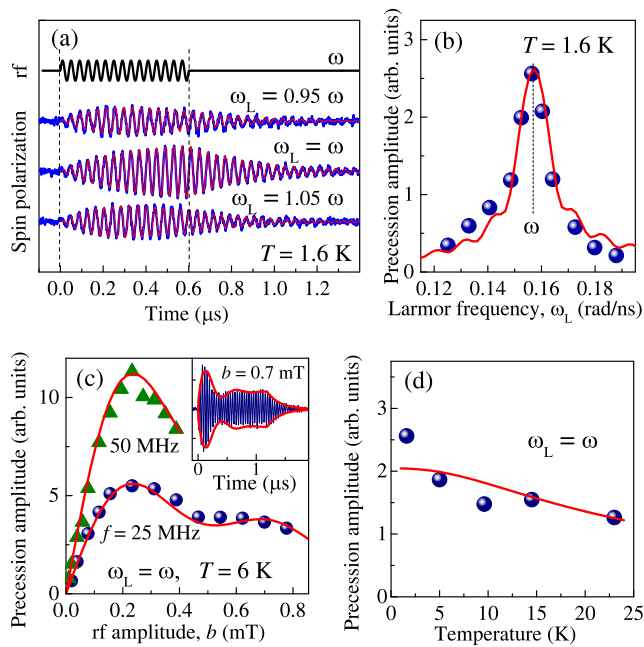}
\caption{(a) Temporal profile of the rf field applied to the $n$-GaAs sample within one pulse and corresponding dynamics of the electron spin polarization (blue lines) for Larmor precession frequencies (controlled by the applied constant magnetic field) varied slightly relative to the rf frequency $f = \omega/2\pi = 25$~MHz. (b),(c),(d) Spin precession amplitude just after the end of the rf pulse as function of the Larmor frequency (magnetic field) (b), of the amplitude of the rf field (c) and of the temperature for $f = 25$~MHz (d). Inset in the panel (c) shows spin dynamics driven by the rf field with high amplitude (0.7~mT), long duration (1.2~$\mu$s) and resonant frequency of 25~MHz. The red lines in all figures show the results of calculations (see text).  
}
\label{fig:wAT}
\end{figure}

\textit{Experimental results.} To examine the technique we apply rf pulses of 15 sinusoidal oscillations with a frequency $f = 25$~MHz [Fig.~\ref{fig:Scheme}(b)] to the GaAs sample. A constant magnetic field $B = 4.2$~mT is applied in the Voigt geometry, so that the electron spin Larmor precession frequency is resonant with that of the rf field, $|g|\mu_\text{B}B/2\pi\hbar = f$. Figure~\ref{fig:Scheme}(c) shows the dynamics of the spin polarization measured as the Faraday rotation of the probe pulse polarization. The spin polarization oscillates at the frequency $f$ with an increasing amplitude due to driving by the rf field. When the rf field is switched off, the spin polarization decays with a characteristic time $T_2 \approx 300$~ns. This decay is similar to the spin dynamics measured by the extended all-optical pump-probe Faraday rotation method, using the same experimental conditions [Fig.~\ref{fig:Scheme}(d)] \cite{Belykh2016}. Interestingly, the increase in the spin precession amplitude with time is determined by $T_2$ following the equation: $1-\exp(-t/T_2)$. We show experimentally and theoretically (see below) that the efficiency of rf spin pumping is determined by frequency and amplitude of the rf field and by the sample temperature.

First, we examine a deviation of the Larmor frequency $\omega_\text{L} = |g|\mu_\text{B}B/\hbar$ from resonance with the rf field frequency $\omega = 2 \pi f$. In these experiments $\omega_\text{L}$ is changed via the magnetic field, while $\omega$ is kept constant. Figure~\ref{fig:wAT}(a) shows the dynamics for $\omega_\text{L} < \omega$, $\omega_\text{L} = \omega$, and $\omega_\text{L} > \omega$. One can see that when $\omega_\text{L}$ is tuned out of resonance, the spin polarization decreases in amplitude and the shape of the dynamics changes. The dependence of the spin precession amplitude just after the rf pulse on the Larmor precession frequency is shown in Fig.~\ref{fig:wAT}(b). It shows a sharp resonance at $\omega = \omega_\text{L}$ with the half width at half maximum determined by the decoherence rate $1/T_2$.

In Fig.~\ref{fig:wAT}(c) we plot the dependence of the spin precession amplitude on the rf field amplitude in the resonant case $\omega = \omega_\text{L}$ for two different frequencies. The amplitude of the rf field $b$ is controlled by the amplitude of the voltage $U$ applied to the rf coil and calculated using the formula $b = U/2\pi^2 f N r^2$, where $N$ is the number of windings and $r$ is the effective radius of the coil. Note, for higher $f$ the maximal rf amplitude that can be achieved is smaller. We take $N = 30$ and $r=0.9$~mm. The dependence in Fig.~\ref{fig:wAT}(c) is linear at low $b$ and for higher $b$ evidences Rabi oscillations \cite{Rabi1937} with the period of $\Delta b = 4\pi\hbar/|g|\mu_\text{B}\mathcal{T}$, where $\mathcal{T}=0.6$~$\mu$s is the duration of the rf pulse. The nonmonotonic behavior of spin polarization related to the Rabi oscillations is also observed in time domain when spin precession is driven by long rf pulse with high amplitude [inset in Fig.~\ref{fig:wAT}(c)]. Note, all data, except those in Fig.~\ref{fig:wAT}(c), are taken for $b \ll \Delta b$.

When the two-level spins system (spin parallel and antiparallel to the magnetic field $\mathbf{B}$) is driven by the electromagnetic field with a frequency corresponding to the level splitting, the response of the system is proportional to the difference of the thermal populations of the levels. Thus, the constant magnetic field (and, correspondingly, the resonance rf frequency) leading to the Zeeman splitting and the temperature $T$ determine the maximal spin precession amplitude that can be reached with rf pumping. Figure~\ref{fig:wAT}(c) shows that for increased frequency from 25 to 50~MHz (and magnetic field from 4.2 to 8.4~mT) the precession amplitude is also increased twice. We also study the dependence of the spin precession amplitude on temperature for $\omega = \omega_\text{L}$ [Fig.~\ref{fig:wAT}(d)]. As it will be shown below [see Eq.~\eqref{eq:Sz}], for $\omega_\text{L} = \omega$ the precession amplitude is proportional to $T_2$ which is also temperature-dependent changing from 300 to 130~ns when $T$ is increased from 1.6 to 23~K. Thus, in Fig.~\ref{fig:wAT}(d) we normalize the precession amplitude to the measured $T_2$. The obtained temperature dependence is rather weak which is related to the degeneracy of the electron gas.

\textit{Model.} The behavior of the total spin polarization $\mathbf{S}$ in a time-dependent magnetic field is described by the Bloch equation \cite{Bloch1946,Abragam1961}:
\begin{equation}
\frac{d\mathbf{S}}{dt} = -\frac{g\mu_\text{B}}{\hbar} \mathbf{S} \times \mathbf{B}_\text{tot} - \hat{\gamma}(\mathbf{S} - \mathbf{S}_\text{st}),
\label{eq:Bloch}
\end{equation}
where $\mathbf{B}_\text{tot} = \mathbf{B} + \mathbf{b}(t)$ is the total magnetic field, $\mathbf{B}$ is its constant component and $\mathbf{b}(t)$ is the time-dependent component induced by the rf coil, $\mathbf{S}_\text{st}$ is the stationary spin polarization in the field $\mathbf{B}_\text{tot}$ that is given by the thermal distribution, and $\hat{\gamma}$ is the relaxation matrix.
In the considered experimental configuration [Fig.~\ref{fig:Scheme}(a)]
\begin{equation}
\mathbf{B}_\text{tot} = \left(\begin{matrix} B\\0\\b\sin(\omega t)\end{matrix}\right),
\end{equation}
$b \ll B$. This corresponds to
\begin{multline}
\hat{\gamma} =
\left(\begin{matrix}
1/T_1 & 0 & 0 \\
0 & 1/T_2 & 0 \\
0 & 0 & 1/T_2 \\
\end{matrix}\right),\\
\mathbf{S}_\text{st} = \left(\begin{matrix} S_\text{st}(B, T, n_0)\\0\\0\end{matrix}\right),
\label{eq:gS1}
\end{multline}
where $T_1$ and $T_2$ are the longitudinal and transverse spin relaxation times, respectively, $S_\text{st}(B, T, n_0)$ is the function giving the total equilibrium spin polarization along the constant field $\mathbf{B}$, which depends on the electron concentration $n_0$ and temperature $T$:
\begin{multline}
S_\text{st}(B, T, n_0) =
\frac{1}{n_0} \int\int S_x dS_x d \Gamma \times\\
\left[\exp\left(\frac{E(k) + g\mu_B B S_x - \mu}{k_\text{B} T}\right)+1\right]^{-1} \\
\approx -\frac{g\mu_B B}{12 n_0}\frac{\partial n(\mu, T)}{\partial \mu}|_{\mu = \mu_0}.
\label{eq:sst}
\end{multline}
Here the second integration $d\Gamma$ is done over the phase space, $\mu$ is the chemical potential (in the limit of $T=0$ it is identical to the Fermi energy), $n(\mu_0, T) = n_0$, we take into account that $|g|\mu_\text{B}B \ll \mu$, and $k_\text{B}$ is the Boltzmann constant.

The results of the $S_z$ calculations for the parameters used in the experiments and for $T_1 = T_2 = 300$~ns are shown in Fig.~\ref{fig:wAT}(a) by the red lines. They accurately reproduce the experimental spin dynamics. The calculated dependence of the spin precession amplitude on the Larmor precession frequency is shown in Fig.~\ref{fig:wAT}(b). The ripples appearing on this dependence originate from the relatively short rf pulse with the local maxima corresponding to integer numbers of Larmor precession periods within the rf pulse width. For longer rf pulses the calculated dependence is similar, but smooth. For the rf pulse being much longer than $T_2$ and for a low rf field amplitude, $|g|\mu_\text{B} b / \hbar \ll 1/T_1,1/T_2$, it follows from the Bloch equations that the spin precession amplitude can be approximated by the following analytical expression:
\begin{equation}
S_z^\text{(a)} \approx \frac{|g|\mu_\text{B} b}{\hbar} \omega_\text{L} \frac{S_\text{st}}{\sqrt{(\omega_\text{L}^2+1/T_2^2-\omega^2)^2+4\omega^2/T_2^2}}.
\label{eq:Sz}
\end{equation}
It, in particular, shows that the spin precession amplitude is proportional to $S_\text{st}$ given by Eq.~\eqref{eq:sst}, which determines the temperature dependence of $S_z^\text{(a)}$.

The model numerical calculations also reproduce the experimental dependencies of the spin precession amplitude on the amplitude of the rf field [Fig.~\ref{fig:wAT}(c)] and on the temperature [Fig.~\ref{fig:wAT}(d)]. We note that in the considered experimental geometry the decay of the dynamics is determined by $T_2$, while the contrast of the Rabi oscillations, which are related to the rf-driven spin transition between the Zeeman-split levels, is determined mainly by the corresponding relaxation time $T_1$.

So far we have considered the precession of a single spin or a homogeneous ensemble of spins with small spread of Larmor precession frequencies $\Delta \omega_\text{L}$ compared to $1/T_2$ (in fact, this is the case for the studied sample with a degenerate electron gas). However, typically in semiconductor structures $\omega_\text{L}$ is rather inhomogeneous within the spin ensemble, and the total spin polarization decays with time $T_2^* \approx 1/\Delta \omega_\text{L} \ll T_2$. On the other hand, the rf excitation with a pulse duration longer than $T_2$ and with $|g|\mu_\text{B} b / \hbar \ll 1/T_1,1/T_2$ resonantly selects a spin subensemble with $\omega_\text{L} = \omega$ within $1/T_2$ interval [Eq.~\eqref{eq:Sz}], and in the case of {\it noninteracting} spins the rise and the decay of spin precession are governed by the time $T_2$. In this case, however, the width of the precession amplitude frequency dependence [Fig.~\ref{fig:wAT}(b)] is determined by the rate $1/T_2^*$.

\begin{figure}
\includegraphics[width=1\columnwidth]{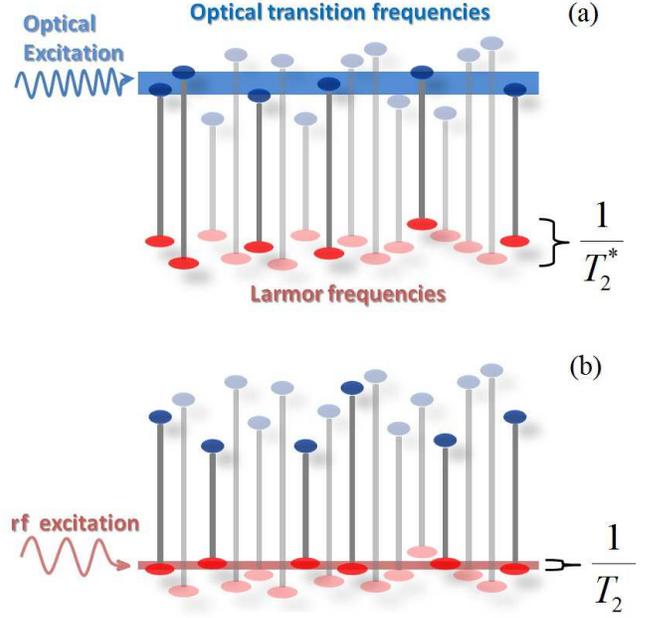}
\caption{System of isolated electron states, each of them characterized by an optical transition frequency and a Larmor frequency. (a) Optical excitation selects electrons with given optical frequencies irrespective of the Larmor frequencies resulting in a large spread of precession frequencies and in spin polarization dephasing with the inhomogeneous time $T_2^*$. (b) rf excitation selects a narrow distribution of Larmor precession frequencies and after sufficiently long and weak rf pulse spin polarization decays with homogeneous time $T_2$.}
\label{fig:T2}
\end{figure}

\textit{Discussion.} We have proposed a new spin-resonant pump-probe method for studying coherent electron spin dynamics based on pumping with radiofrequency field and optical probing of the Faraday (or Kerr) rotation. 
The developed technique is scalable up to GHz-range frequencies, corresponding to the Tesla-range of magnetic fields. Here the limitation will be the jitter between the rf and optical pulses which should be smaller than $1/\omega$. With up to date generators this jitter can be as small as 25~ps, corresponding to a maximal frequency of 6~GHz and a magnetic field of 1~T (for $n$-GaAs). The technique is simpler to realize than all-optical pump-probe Faraday rotation: it is a single-beam technique (only probe) and does not require a mechanical delay line and a helicity modulator for the pump, a precise adjustment of the pump and probe beam coincidence on the sample. Furthermore, in the all-optical pump-probe Faraday rotation scheme, the pump addresses optical transitions with a given optical frequency orienting spins irrespective of their Larmor frequencies whose distribution has a width $1/T_2^*$ [Fig.~\ref{fig:T2}(a)]. So the spin polarization decays with time $T_2^*$. In the developed technique, the rf pump directly addresses the electron spins whose Larmor frequencies are close to the rf frequency within the $1/T_2$ interval (for a rf pulse width longer than $T_2$ and for a not too large rf amplitude), where the spin coherence time $T_2$ corresponds to that of an individual spin rather than to the spin ensemble [Fig.~\ref{fig:T2}(b)]. As a result, the spin polarization decays with time $T_2$.  In the studied $n$-GaAs sample the electron density exceeds the the metal-insulator transition and the majority of the electrons is free. Therefore, the individual spin properties are shared over the spin ensemble due to spin exchange averaging \cite{Pines1955,Paget1981}, washing out the inhomogeneity and leading to $T_2 \approx T_2^*$. However, in inhomogeneous systems with strongly localized electrons such as $n$-GaAs with a low donor concentration or quantum dots, $T_2 \gg T_2^*$ \cite{Greilich2006} and the developed technique provides a simple way of measuring $T_2$, while all-optical pump-probe methods give access to the ensemble-related, much shorter inhomogeneous spin dephasing time $T_2^*$.

\begin{acknowledgments}
\textit{Acknowledgments.} We are grateful to I.~A.~Akimov and D.~O.~Tolmachev for valuable advices and discussions. We acknowledge the financial support of the Russian Science Foundation through the grant No.18-72-10073 (experiments with high rf amplitudes and model) and the Deutsche Forschungsgemeinschaft (DFG) in the frame of the ICRC TRR 160 (project A1).
\end{acknowledgments}


\begin{thebibliography}{20}%
\makeatletter
\providecommand \@ifxundefined [1]{%
 \@ifx{#1\undefined}
}%
\providecommand \@ifnum [1]{%
 \ifnum #1\expandafter \@firstoftwo
 \else \expandafter \@secondoftwo
 \fi
}%
\providecommand \@ifx [1]{%
 \ifx #1\expandafter \@firstoftwo
 \else \expandafter \@secondoftwo
 \fi
}%
\providecommand \natexlab [1]{#1}%
\providecommand \enquote  [1]{``#1''}%
\providecommand \bibnamefont  [1]{#1}%
\providecommand \bibfnamefont [1]{#1}%
\providecommand \citenamefont [1]{#1}%
\providecommand \href@noop [0]{\@secondoftwo}%
\providecommand \href [0]{\begingroup \@sanitize@url \@href}%
\providecommand \@href[1]{\@@startlink{#1}\@@href}%
\providecommand \@@href[1]{\endgroup#1\@@endlink}%
\providecommand \@sanitize@url [0]{\catcode `\\12\catcode `\$12\catcode
  `\&12\catcode `\#12\catcode `\^12\catcode `\_12\catcode `\%12\relax}%
\providecommand \@@startlink[1]{}%
\providecommand \@@endlink[0]{}%
\providecommand \url  [0]{\begingroup\@sanitize@url \@url }%
\providecommand \@url [1]{\endgroup\@href {#1}{\urlprefix }}%
\providecommand \urlprefix  [0]{URL }%
\providecommand \Eprint [0]{\href }%
\providecommand \doibase [0]{http://dx.doi.org/}%
\providecommand \selectlanguage [0]{\@gobble}%
\providecommand \bibinfo  [0]{\@secondoftwo}%
\providecommand \bibfield  [0]{\@secondoftwo}%
\providecommand \translation [1]{[#1]}%
\providecommand \BibitemOpen [0]{}%
\providecommand \bibitemStop [0]{}%
\providecommand \bibitemNoStop [0]{.\EOS\space}%
\providecommand \EOS [0]{\spacefactor3000\relax}%
\providecommand \BibitemShut  [1]{\csname bibitem#1\endcsname}%
\let\auto@bib@innerbib\@empty
\bibitem [{\citenamefont {Baumberg}\ \emph {et~al.}(1994)\citenamefont
  {Baumberg}, \citenamefont {Awschalom}, \citenamefont {Samarth}, \citenamefont
  {Luo},\ and\ \citenamefont {Furdyna}}]{Baumberg1994}%
  \BibitemOpen
  \bibfield  {author} {\bibinfo {author} {\bibfnamefont {J.~J.}\ \bibnamefont
  {Baumberg}}, \bibinfo {author} {\bibfnamefont {D.~D.}\ \bibnamefont
  {Awschalom}}, \bibinfo {author} {\bibfnamefont {N.}~\bibnamefont {Samarth}},
  \bibinfo {author} {\bibfnamefont {H.}~\bibnamefont {Luo}}, \ and\ \bibinfo
  {author} {\bibfnamefont {J.~K.}\ \bibnamefont {Furdyna}},\ }\bibfield
  {title} {\enquote {\bibinfo {title} {{Spin beats and dynamical magnetization
  in quantum structures}},}\ }\href {\doibase 10.1103/PhysRevLett.72.717}
  {\bibfield  {journal} {\bibinfo  {journal} {Phys. Rev. Lett.}\ }\textbf
  {\bibinfo {volume} {72}},\ \bibinfo {pages} {717} (\bibinfo {year}
  {1994})}\BibitemShut {NoStop}%
\bibitem [{\citenamefont {Zheludev}\ \emph {et~al.}(1994)\citenamefont
  {Zheludev}, \citenamefont {Brummell}, \citenamefont {Harley}, \citenamefont
  {Malinowski}, \citenamefont {Popov}, \citenamefont {Ashenford},\ and\
  \citenamefont {Lunn}}]{Zheludev1994}%
  \BibitemOpen
  \bibfield  {author} {\bibinfo {author} {\bibfnamefont {N.I.}\ \bibnamefont
  {Zheludev}}, \bibinfo {author} {\bibfnamefont {M.A.}\ \bibnamefont
  {Brummell}}, \bibinfo {author} {\bibfnamefont {R.T.}\ \bibnamefont {Harley}},
  \bibinfo {author} {\bibfnamefont {A.}~\bibnamefont {Malinowski}}, \bibinfo
  {author} {\bibfnamefont {S.V.}\ \bibnamefont {Popov}}, \bibinfo {author}
  {\bibfnamefont {D.E.}\ \bibnamefont {Ashenford}}, \ and\ \bibinfo {author}
  {\bibfnamefont {B.}~\bibnamefont {Lunn}},\ }\bibfield  {title} {\enquote
  {\bibinfo {title} {{Giant specular inverse Faraday effect in
  Cd0.6Mn0.4Te}},}\ }\href {\doibase 10.1016/0038-1098(94)90064-7} {\bibfield
  {journal} {\bibinfo  {journal} {Solid State Commun.}\ }\textbf {\bibinfo
  {volume} {89}},\ \bibinfo {pages} {823} (\bibinfo {year} {1994})}\BibitemShut
  {NoStop}%
\bibitem [{\citenamefont {Yakovlev}\ and\ \citenamefont
  {Bayer}(2017)}]{Dyakonov2017}%
  \BibitemOpen
  \bibfield  {author} {\bibinfo {author} {\bibfnamefont {D.R.}\ \bibnamefont
  {Yakovlev}}\ and\ \bibinfo {author} {\bibfnamefont {M.}~\bibnamefont
  {Bayer}},\ }\bibfield  {title} {\enquote {\bibinfo {title} {{Coherent Spin
  Dynamics of Carriers}},}\ }in\ \href {\doibase 10.1007/978-3-319-65436-2_6}
  {\emph {\bibinfo {booktitle} {Spin Phys. Semicond.}}},\ \bibinfo {series}
  {Springer Series in Solid-State Sciences}, Vol.\ \bibinfo {volume} {157},\
  \bibinfo {editor} {edited by\ \bibinfo {editor} {\bibfnamefont {Mikhail~I.}\
  \bibnamefont {Dyakonov}}}\ (\bibinfo  {publisher} {Springer International
  Publishing},\ \bibinfo {address} {Cham},\ \bibinfo {year} {2017})\ p.\
  \bibinfo {pages} {155}\BibitemShut {NoStop}%
\bibitem [{\citenamefont {Belykh}\ \emph {et~al.}(2016)\citenamefont {Belykh},
  \citenamefont {Evers}, \citenamefont {Yakovlev}, \citenamefont {Fobbe},
  \citenamefont {Greilich},\ and\ \citenamefont {Bayer}}]{Belykh2016}%
  \BibitemOpen
  \bibfield  {author} {\bibinfo {author} {\bibfnamefont {V.~V.}\ \bibnamefont
  {Belykh}}, \bibinfo {author} {\bibfnamefont {E.}~\bibnamefont {Evers}},
  \bibinfo {author} {\bibfnamefont {D.~R.}\ \bibnamefont {Yakovlev}}, \bibinfo
  {author} {\bibfnamefont {F.}~\bibnamefont {Fobbe}}, \bibinfo {author}
  {\bibfnamefont {A.}~\bibnamefont {Greilich}}, \ and\ \bibinfo {author}
  {\bibfnamefont {M.}~\bibnamefont {Bayer}},\ }\bibfield  {title} {\enquote
  {\bibinfo {title} {{Extended pump-probe Faraday rotation spectroscopy of the
  submicrosecond electron spin dynamics in n-type GaAs}},}\ }\href {\doibase
  10.1103/PhysRevB.94.241202} {\bibfield  {journal} {\bibinfo  {journal} {Phys.
  Rev. B}\ }\textbf {\bibinfo {volume} {94}},\ \bibinfo {pages} {241202(R)}
  (\bibinfo {year} {2016})}\BibitemShut {NoStop}%
\bibitem [{\citenamefont {Aleksandrov}\ and\ \citenamefont
  {Zapasskii}(1981)}]{Aleksandrov1981}%
  \BibitemOpen
  \bibfield  {author} {\bibinfo {author} {\bibfnamefont {E.~B.}\ \bibnamefont
  {Aleksandrov}}\ and\ \bibinfo {author} {\bibfnamefont {V.~S.}\ \bibnamefont
  {Zapasskii}},\ }\bibfield  {title} {\enquote {\bibinfo {title} {{Magnetic
  resonance in the Faraday-rotation noise spectrum}},}\ }\href@noop {}
  {\bibfield  {journal} {\bibinfo  {journal} {Sov. Phys. JETP}\ }\textbf
  {\bibinfo {volume} {54}},\ \bibinfo {pages} {64} (\bibinfo {year}
  {1981})}\BibitemShut {NoStop}%
\bibitem [{\citenamefont {Crooker}\ \emph {et~al.}(2004)\citenamefont
  {Crooker}, \citenamefont {Rickel}, \citenamefont {Balatsky},\ and\
  \citenamefont {Smith}}]{Crooker2004}%
  \BibitemOpen
  \bibfield  {author} {\bibinfo {author} {\bibfnamefont {S.~A.}\ \bibnamefont
  {Crooker}}, \bibinfo {author} {\bibfnamefont {D.~G.}\ \bibnamefont {Rickel}},
  \bibinfo {author} {\bibfnamefont {A.~V.}\ \bibnamefont {Balatsky}}, \ and\
  \bibinfo {author} {\bibfnamefont {D.~L.}\ \bibnamefont {Smith}},\ }\bibfield
  {title} {\enquote {\bibinfo {title} {{Spectroscopy of spontaneous spin noise
  as a probe of spin dynamics and magnetic resonance}},}\ }\href {\doibase
  10.1038/nature02804} {\bibfield  {journal} {\bibinfo  {journal} {Nature}\
  }\textbf {\bibinfo {volume} {431}},\ \bibinfo {pages} {49} (\bibinfo {year}
  {2004})}\BibitemShut {NoStop}%
\bibitem [{\citenamefont {Oestreich}\ \emph {et~al.}(2005)\citenamefont
  {Oestreich}, \citenamefont {R{\"{o}}mer}, \citenamefont {Haug},\ and\
  \citenamefont {H{\"{a}}gele}}]{Oestreich2005}%
  \BibitemOpen
  \bibfield  {author} {\bibinfo {author} {\bibfnamefont {M.}~\bibnamefont
  {Oestreich}}, \bibinfo {author} {\bibfnamefont {M.}~\bibnamefont
  {R{\"{o}}mer}}, \bibinfo {author} {\bibfnamefont {R.~J.}\ \bibnamefont
  {Haug}}, \ and\ \bibinfo {author} {\bibfnamefont {D.}~\bibnamefont
  {H{\"{a}}gele}},\ }\bibfield  {title} {\enquote {\bibinfo {title} {{Spin
  Noise Spectroscopy in GaAs}},}\ }\href {\doibase
  10.1103/PhysRevLett.95.216603} {\bibfield  {journal} {\bibinfo  {journal}
  {Phys. Rev. Lett.}\ }\textbf {\bibinfo {volume} {95}},\ \bibinfo {pages}
  {216603} (\bibinfo {year} {2005})}\BibitemShut {NoStop}%
\bibitem [{\citenamefont {Blume}(1958)}]{Blume1958}%
  \BibitemOpen
  \bibfield  {author} {\bibinfo {author} {\bibfnamefont {R.~J.}\ \bibnamefont
  {Blume}},\ }\bibfield  {title} {\enquote {\bibinfo {title} {{Electron Spin
  Relaxation Times in Sodium-Ammonia Solutions}},}\ }\href {\doibase
  10.1103/PhysRev.109.1867} {\bibfield  {journal} {\bibinfo  {journal} {Phys.
  Rev.}\ }\textbf {\bibinfo {volume} {109}},\ \bibinfo {pages} {1867} (\bibinfo
  {year} {1958})}\BibitemShut {NoStop}%
\bibitem [{\citenamefont {Gordon}\ and\ \citenamefont
  {Bowers}(1958)}]{Gordon1958}%
  \BibitemOpen
  \bibfield  {author} {\bibinfo {author} {\bibfnamefont {J.~P.}\ \bibnamefont
  {Gordon}}\ and\ \bibinfo {author} {\bibfnamefont {K.~D.}\ \bibnamefont
  {Bowers}},\ }\bibfield  {title} {\enquote {\bibinfo {title} {{Microwave Spin
  Echoes from Donor Electrons in Silicon}},}\ }\href {\doibase
  10.1103/PhysRevLett.1.368} {\bibfield  {journal} {\bibinfo  {journal} {Phys.
  Rev. Lett.}\ }\textbf {\bibinfo {volume} {1}},\ \bibinfo {pages} {368}
  (\bibinfo {year} {1958})}\BibitemShut {NoStop}%
\bibitem [{\citenamefont {Schweiger}\ and\ \citenamefont
  {Jeschke}(2001)}]{Schweiger2001}%
  \BibitemOpen
  \bibfield  {author} {\bibinfo {author} {\bibfnamefont {A.}~\bibnamefont
  {Schweiger}}\ and\ \bibinfo {author} {\bibfnamefont {G.}~\bibnamefont
  {Jeschke}},\ }\href@noop {} {\emph {\bibinfo {title} {{Principles of pulse
  electron paramagnetic resonance}}}}\ (\bibinfo  {publisher} {Oxford
  University Press},\ \bibinfo {address} {New York},\ \bibinfo {year}
  {2001})\BibitemShut {NoStop}%
\bibitem [{\citenamefont {Dawson}\ and\ \citenamefont
  {Cavenett}(1979)}]{Dawson1979}%
  \BibitemOpen
  \bibfield  {author} {\bibinfo {author} {\bibfnamefont {P.}~\bibnamefont
  {Dawson}}\ and\ \bibinfo {author} {\bibfnamefont {B.C.}\ \bibnamefont
  {Cavenett}},\ }\bibfield  {title} {\enquote {\bibinfo {title} {{Time resolved
  optically detected resonance in ZnS}},}\ }\href {\doibase
  10.1016/0022-2313(79)90250-3} {\bibfield  {journal} {\bibinfo  {journal} {J.
  Lumin.}\ }\textbf {\bibinfo {volume} {18-19}},\ \bibinfo {pages} {853}
  (\bibinfo {year} {1979})}\BibitemShut {NoStop}%
\bibitem [{\citenamefont {Trifunac}\ and\ \citenamefont
  {Smith}(1980)}]{Trifunac1980}%
  \BibitemOpen
  \bibfield  {author} {\bibinfo {author} {\bibfnamefont {A.~D.}\ \bibnamefont
  {Trifunac}}\ and\ \bibinfo {author} {\bibfnamefont {J.~P.}\ \bibnamefont
  {Smith}},\ }\bibfield  {title} {\enquote {\bibinfo {title} {{Optically
  detected time resolved EPR of radical ion pairs in pulse radiolysis of
  liquids}},}\ }\href {\doibase 10.1016/0009-2614(80)85210-9} {\bibfield
  {journal} {\bibinfo  {journal} {Chem. Phys. Lett.}\ }\textbf {\bibinfo
  {volume} {73}},\ \bibinfo {pages} {94} (\bibinfo {year} {1980})}\BibitemShut
  {NoStop}%
\bibitem [{\citenamefont {Mlynek}\ \emph {et~al.}(1983)\citenamefont {Mlynek},
  \citenamefont {Wong}, \citenamefont {DeVoe}, \citenamefont {Kintzer},\ and\
  \citenamefont {Brewer}}]{Mlynek1983}%
  \BibitemOpen
  \bibfield  {author} {\bibinfo {author} {\bibfnamefont {J.}~\bibnamefont
  {Mlynek}}, \bibinfo {author} {\bibfnamefont {N.~C.}\ \bibnamefont {Wong}},
  \bibinfo {author} {\bibfnamefont {R.~G.}\ \bibnamefont {DeVoe}}, \bibinfo
  {author} {\bibfnamefont {E.~S.}\ \bibnamefont {Kintzer}}, \ and\ \bibinfo
  {author} {\bibfnamefont {R.~G.}\ \bibnamefont {Brewer}},\ }\bibfield  {title}
  {\enquote {\bibinfo {title} {{Raman Heterodyne Detection of Nuclear Magnetic
  Resonance}},}\ }\href {\doibase 10.1103/PhysRevLett.50.993} {\bibfield
  {journal} {\bibinfo  {journal} {Phys. Rev. Lett.}\ }\textbf {\bibinfo
  {volume} {50}},\ \bibinfo {pages} {993} (\bibinfo {year} {1983})}\BibitemShut
  {NoStop}%
\bibitem [{\citenamefont {Wei}\ \emph {et~al.}(1996)\citenamefont {Wei},
  \citenamefont {Holmstrom}, \citenamefont {Manson}, \citenamefont {Martin},
  \citenamefont {He}, \citenamefont {Fisk},\ and\ \citenamefont
  {Holliday}}]{Wei1996}%
  \BibitemOpen
  \bibfield  {author} {\bibinfo {author} {\bibfnamefont {C.}~\bibnamefont
  {Wei}}, \bibinfo {author} {\bibfnamefont {S.~A.}\ \bibnamefont {Holmstrom}},
  \bibinfo {author} {\bibfnamefont {N.~B.}\ \bibnamefont {Manson}}, \bibinfo
  {author} {\bibfnamefont {J.~P.~D.}\ \bibnamefont {Martin}}, \bibinfo {author}
  {\bibfnamefont {X.~F.}\ \bibnamefont {He}}, \bibinfo {author} {\bibfnamefont
  {P.~T.~H.}\ \bibnamefont {Fisk}}, \ and\ \bibinfo {author} {\bibfnamefont
  {K.}~\bibnamefont {Holliday}},\ }\bibfield  {title} {\enquote {\bibinfo
  {title} {{Raman heterodyne detected magnetic resonance: I. CW and coherent
  transient measurements}},}\ }\href {\doibase 10.1007/BF03162247} {\bibfield
  {journal} {\bibinfo  {journal} {Appl. Magn. Reson.}\ }\textbf {\bibinfo
  {volume} {11}},\ \bibinfo {pages} {521} (\bibinfo {year} {1996})}\BibitemShut
  {NoStop}%
\bibitem [{\citenamefont {Bloch}(1946)}]{Bloch1946}%
  \BibitemOpen
  \bibfield  {author} {\bibinfo {author} {\bibfnamefont {F.}~\bibnamefont
  {Bloch}},\ }\bibfield  {title} {\enquote {\bibinfo {title} {{Nuclear
  Induction}},}\ }\href {\doibase 10.1103/PhysRev.70.460} {\bibfield  {journal}
  {\bibinfo  {journal} {Phys. Rev.}\ }\textbf {\bibinfo {volume} {70}},\
  \bibinfo {pages} {460} (\bibinfo {year} {1946})}\BibitemShut {NoStop}%
\bibitem [{\citenamefont {Abragam}(1961)}]{Abragam1961}%
  \BibitemOpen
  \bibfield  {author} {\bibinfo {author} {\bibfnamefont {A.}~\bibnamefont
  {Abragam}},\ }\href@noop {} {\emph {\bibinfo {title} {{Principles of Nuclear
  Magnetism}}}}\ (\bibinfo  {publisher} {Clarendon Press},\ \bibinfo {year}
  {1961})\BibitemShut {NoStop}%
\bibitem [{\citenamefont {Rabi}(1937)}]{Rabi1937}%
  \BibitemOpen
  \bibfield  {author} {\bibinfo {author} {\bibfnamefont {I.~I.}\ \bibnamefont
  {Rabi}},\ }\bibfield  {title} {\enquote {\bibinfo {title} {{Space
  Quantization in a Gyrating Magnetic Field}},}\ }\href {\doibase
  10.1103/PhysRev.51.652} {\bibfield  {journal} {\bibinfo  {journal} {Phys.
  Rev.}\ }\textbf {\bibinfo {volume} {51}},\ \bibinfo {pages} {652} (\bibinfo
  {year} {1937})}\BibitemShut {NoStop}%
\bibitem [{\citenamefont {Pines}\ and\ \citenamefont
  {Slichter}(1955)}]{Pines1955}%
  \BibitemOpen
  \bibfield  {author} {\bibinfo {author} {\bibfnamefont {D.}~\bibnamefont
  {Pines}}\ and\ \bibinfo {author} {\bibfnamefont {C.~P.}\ \bibnamefont
  {Slichter}},\ }\bibfield  {title} {\enquote {\bibinfo {title} {{Relaxation
  Times in Magnetic Resonance}},}\ }\href {\doibase 10.1103/PhysRev.100.1014}
  {\bibfield  {journal} {\bibinfo  {journal} {Phys. Rev.}\ }\textbf {\bibinfo
  {volume} {100}},\ \bibinfo {pages} {1014} (\bibinfo {year}
  {1955})}\BibitemShut {NoStop}%
\bibitem [{\citenamefont {Paget}(1981)}]{Paget1981}%
  \BibitemOpen
  \bibfield  {author} {\bibinfo {author} {\bibfnamefont {D.}~\bibnamefont
  {Paget}},\ }\bibfield  {title} {\enquote {\bibinfo {title} {{Optical
  detection of NMR in high-purity GaAs under optical pumping: Efficient
  spin-exchange averaging between electronic states}},}\ }\href {\doibase
  10.1103/PhysRevB.24.3776} {\bibfield  {journal} {\bibinfo  {journal} {Phys.
  Rev. B}\ }\textbf {\bibinfo {volume} {24}},\ \bibinfo {pages} {3776}
  (\bibinfo {year} {1981})}\BibitemShut {NoStop}%
\bibitem [{\citenamefont {Greilich}\ \emph {et~al.}(2006)\citenamefont
  {Greilich}, \citenamefont {Yakovlev}, \citenamefont {Shabaev}, \citenamefont
  {Efros}, \citenamefont {Yugova}, \citenamefont {Oulton}, \citenamefont
  {Stavarache}, \citenamefont {Reuter}, \citenamefont {Wieck},\ and\
  \citenamefont {Bayer}}]{Greilich2006}%
  \BibitemOpen
  \bibfield  {author} {\bibinfo {author} {\bibfnamefont {A.}~\bibnamefont
  {Greilich}}, \bibinfo {author} {\bibfnamefont {D.~R.}\ \bibnamefont
  {Yakovlev}}, \bibinfo {author} {\bibfnamefont {A.}~\bibnamefont {Shabaev}},
  \bibinfo {author} {\bibfnamefont {Al.~L.}\ \bibnamefont {Efros}}, \bibinfo
  {author} {\bibfnamefont {I.~A.}\ \bibnamefont {Yugova}}, \bibinfo {author}
  {\bibfnamefont {R.}~\bibnamefont {Oulton}}, \bibinfo {author} {\bibfnamefont
  {V.}~\bibnamefont {Stavarache}}, \bibinfo {author} {\bibfnamefont
  {D.}~\bibnamefont {Reuter}}, \bibinfo {author} {\bibfnamefont
  {A.}~\bibnamefont {Wieck}}, \ and\ \bibinfo {author} {\bibfnamefont
  {M.}~\bibnamefont {Bayer}},\ }\bibfield  {title} {\enquote {\bibinfo {title}
  {{Mode locking of electron spin coherences in singly charged quantum
  dots}},}\ }\href {\doibase 10.1126/science.1128215} {\bibfield  {journal}
  {\bibinfo  {journal} {Science}\ }\textbf {\bibinfo {volume} {313}},\ \bibinfo
  {pages} {341} (\bibinfo {year} {2006})}\BibitemShut {NoStop}%
\end{thebibliography}
\end{document}